\documentclass[12pt,twoside,reqno]{amspaper}
\usepackage[edges]{mathdef}
\usepackage{natbib}
\bibpunct{(}{)}{,}{a}{}{,}

\usepackage{pstricks,pst-node}
\usepackage{psfrag}
\usepackage{comment}
\begin{document}
\title{A note on global Markov properties for mixed graphs}
\author{Michael Eichler}
\affil{Maastricht University}
\thanks{{\em Current address:} Department of Quantitative Economics,
Maastricht University, P.O.~Box 616, 6200 MD Maastricht, The Netherlands}
\thanks{{\em E-mail address:} m.eichler@maastrichtuniversity.nl (M.~Eichler)}
\dedicatory{{\upshape\today}}
\begin{abstract}
Global Markov properties in mixed graphs are usually formulated in terms of the path-oriented $m$-separation or by use of augmented graphs (similar to moral graphs in the case of directed acyclic graphs). We provide an alternative characterization that can be easily implemented.\\[1em]
{{\itshape Keywords:} Graphical models, separation, global Markov property}
\end{abstract}
\maketitle

\section{Graphical terminology}

The graphs that are used in this paper are mixed graphs with possibly
two kind of edges, namely directed and bi-directed edges. Suppose that
$V$ is a finite and nonempty set. Then a {\em graph} $G$ over $V$ is given
by an ordered pair $(V,E)$ where the elements in $V$ represent the
{\em vertices} or {\em nodes} of the graph and $E$ is a collection of
{\em edges} $e$ denoted as $a\DE b$, $a\LDE b$, or $a\BE b$ for distinct nodes $a,b$ in $V$. The edges $a\DE b$ and $a\LDE b$ are called {\em directed edges} while $a\BE b$ is called a {\em bi-directed edge}\footnote{In \citet{eichlerpathdiagr} mixed graphs with dashed undirected edges $a\UD b$ in place of bi-directed edges $a\BE b$ are considered. The results of this paper apply also to these graphs with the obvious changes in notation.} If $e=a\DE b$, then $e$ has an {\em arrowhead} at $b$ and a {\em tail} at $a$. Similarly, if $e$ is a bi-directed edge $a\BE b$, then $e$ has an arrowhead at both ends $a$ and $b$.

Two nodes $a$ and $b$ that are connected by an edge in $G$ are said to be {\em adjacent} in $G$. If the edge is bi-directed, the two nodes $a$ and $b$ are said to be {\em spouses}. If $a\DE b\in E$ then $a$ is a {\em parent} of $b$ and $b$ is a {\em child} of $a$. The sets of all spouses, parents, and children of $a$ are denoted by $\spouse{G}{a}$, $\parentin{G}{a}$, and $\childin{G}{a}$, respectively. If it is clear which graph $G$ is meant we omit the index $G$. Furthermore, for a subset $A$ of $V$, let $\spouse{A}$, $\parent{A}$, and $\child{A}$ denote the collection of neighbours, parents, and children, respectively, of vertices in $A$ that are not themselves elements of $A$, that is, $\parent{A}=\cup_{a\in A}\parent{a}\without A$ etc. Furthermore, the district of a vertex $a$ is the set of all vertices $b\in V$ that are connected to $a$ by an path $b\BE\ldots\BE a$.

As in \citet{frydenberg}, a node $b$ is said to be an {\em ancestor} of $a$ if
either $b=a$ or there exists a directed path $b\DE\cdots\DE a$ in $G$. The set of all ancestors of elements in $A$ is denoted by $\ancestor{A}$. Notice that this definition differs from the one given in \citet{SL96}, where the vertex $a$ itself is not contained in the set of ancestors. Furthermore, we say that a subset $A$ is {\em ancestral} if it contains all its ancestors, that is, $\ancestor{A}=A$.

Finally, let $G=(V,E)$ and $G'=(V',E')$ be mixed graphs. Then $G'$ is a {\em subgraph} of $G$ if $V'\subseteq V$ and $E'\subseteq E$. If $A$ is a subset of $V$ it induces the subgraph $G_A=(A,E_A)$ where $E_A$ contains all edges $e\in E$ that have both endpoints in $A$.

\section{Separation in mixed graphs}

There are two commonly used criteria for separation in general mixed graphs: the {\em $m$-separation criterion}, which is path-oriented, and the {\em augmentation separation criterion}, which utilizes ordinary separation in undirected graphs.

A {\em path} $\pi$ between two vertices $a$ and $b$ in $G$ is a sequence
$\pi=\path{e_1,\ldots,e_n}$ of edges $e_i\in E$ such that $e_i$ is an
edge between $v_{i-1}$ and $v_i$ for some sequence of vertices
$v_0=a,v_1,\ldots,v_{n}=b$. We say that $a$ and $b$ are the endpoints of
the path, while $v_1,\ldots,v_{n-1}$ are the {\em intermediate vertices}
on the path. Note that the vertices $v_i$ in the sequence do not need to
be distinct and that therefore paths may be self-intersecting.

An intermediate vertex $c$ on a path $\pi$ is said to be a
{\em collider} on the path if the edges preceding and suceeding
$c$ on the path both have arrowheads at $c$,
i.e.~$\DE c \LDE$, $\BE c \BE$, $\BE c \LDE$, $\DE c \BE$;
otherwise the vertex $c$ is said to be a {\em non-collider} on the path\footnote{In the case of graphs with dashed undirected edges $a\UD b$, a dashed tail is viewed as having an arrowhead to apply the definition of colliders and non-colliders.} A path $\pi$ between vertices $a$ and $b$ is said to be {\em $m$-connecting}\footnote{We note that condition (ii) differs from
the original definition of $m$-connecting paths given in \citet{TR03}.
Our simpler condition accounts for the fact that we consider paths
that may be self-intersecting \citep[for a similar definition see][]{koster02}. Despite the difference, the concepts of $m$-separations here and in \citet{TR03}
are equivalent.} given a set $C$ if
\begin{romanlist}
\item
every non-collider on the path is not in $C$, and
\item
every collider on the path is in $C$,
\end{romanlist}
otherwise we say the path is {\em $m$-blocked} given $C$.
If all paths between $a$ and $b$ are $m$-blocked given $C$, then
$a$ and $b$ are said to be {\em $m$-separated} given $C$.
Similarly, sets $A$ and $B$ are said to be $m$-separated in $G$
given $C$, denoted by $A\sepm B\given C\wrt{G}$ if for every
pair $a\in A$ and  $b\in B$, $a$ and $b$ are $m$-separated given $C$.

The augmentation separation criterion in mixed graphs is based on the notion of {\em pure collider paths}, which are defined as paths on which every
intermediate vertex is a collider. Then two vertices $a$ and $b$ are said to be {\em collider connected} if they are connected by a pure collider path. Since every single edge trivially forms a collider path, any two vertices adjacent in $G$ are collider connected.

The {\em augmented graph} $G\augment=(V,E\augment)$ derived from $G$
is an undirected graph with the same vertex set as $G$ and
undirected edges satisfying
\[
a\UE b\in E\augment\iff\text{$a$ and $b$ are collider connected in $G$}.
\]
Let $A$, $B$, and $S$ be disjoint subsets of $V$. We say that $C$ {\em separates} $A$ and $B$ in $G\augment$, denoted by $A\sep B\given C\wrt{G\augment}$, if every path $a\UE\cdots\UE b$ in $G\augment$ between vertices $a\in A$ and $b\in B$ intersects $C$.

\section{An alternative characterization of separation in mixed graphs}

In order to establish that two sets $A$ and $B$ are $m$-separated given a third set $C$, we must show that there does not exist a path between $A$ and $B$ that is $m$-connecting given $C$. As paths are allowed to be self-intersecting, the number of paths between $A$ and $B$ is infinite. Although the search for $m$-connecting paths can be restricted to paths where no edges occurs twice with the same orientation \citep[cf][]{eichlergraphmodel}, an algorithmic implementation of such a search seems not straightforward. In the following, we present an alternative characterization of $m$-separation that is based on an enlargement of the two sets $A$ and $B$.

\begin{theorem}
\label{theorem}
Let $G=(V,E)$ be a mixed graph and let $A$, $B$, and $C$ be three disjoint subsets of $V$. Then the following are equivalent:
\begin{romanlist}
\item
$A\sepm B\given C\wrt{G}$
\item
$A\sep B\given C\wrt{(G_{\ancestor{A\cup B\cup C}})^\textrm{a}}$
\item
there exist two disjoint subsets $A^*$ and $B^*$ such that $A\subseteq A^*$, $B\subseteq B^*$, $V^*=A^*\cup B^*\cup C=\ancestor{A\cup B\cup C}$ and
\[
\districtin{G^*}{A^*\cup\child{A^*}}\cap\districtin{G^*}{B^*\cup\child{B^*}}
=\varempty,
\]
where $G^*=G_{V^*}$ is the subgraph of $G$ induced by the subset $V^*$.
\end{romanlist}
\end{theorem}

The proof of the theorem is based  on the following lemma.

\begin{lemma}
Let $G=(V,E)$ be a mixed graph, and let $A$ and  $B$ be two disjoint subsets of $V$. Then the following statements are equivalent:
\begin{romanlist}
\item
$A\sepm B\given V\without(A\cup B)$;
\item
$A$ and $B$ are not connected by some pure-collider path;
\item
$\district{A\cup\child{A}}\cap\district{B\cup\child{B}}=\varempty$.
\end{romanlist}
\end{lemma}
\begin{proof}
From the definition of $m$-separation it follows that a path between
$a$ and $b$ with all intermediate vertices not in $A$ or $B$ is $m$-connecting given $V\without(A\cup B)$ if and only if all intermediate vertices on the path are $m$-colliders and hence the path is a pure-collider path. Since a vertex $v$ is an $m$-collider if and only if none of the two adjacent edges is directed with its tail at $v$, a pure-collider path between vertices $a$ and $b$ is necessarily of  the form
\begin{romanlist}
\item
$a\BE\cdots\BE b$;
\item
$a\DE c\BE\cdots\BE b$;
\item
$a\BE\cdots\BE c\LDE b$;
\item
$a\DE c\BE\cdots\BE d\LDE b$.
\end{romanlist}
Now suppose that two vertices $a\in A$ and $b\in B$ are $m$-connected given $V\without(A\cup B)$, and let $\pi$ be the corresponding $m$-connecting path.
Then there exists a subpath $\pi'$ between vertices $a'\in A$ and $b'\in B$ such that every intermediate vertex on $\pi'$ is in $V\without(A\cup B)$. By the arguments above it follows that $\pi'$ is a pure-collider path and thus is of one of the types (i) to (iv). Conversely, if $\pi$ is a pure-collider path between $a$ and $b$, then $\pi$ has a subpath $\pi'$ between vertices $a'\in A$ and $b'\in B$ such that all intermediate vertices are neither in $A$ nor in $B$. This implies that $\pi'$ is $m$-connecting given $V\without(A\cup B)$. This shows the equivalence of (i) and (ii).

Next, for the equivalence of conditions (ii) and (iii), we note that for the four types of pure-collider pathes between $a$ and $b$ we have
\begin{alphlist}
\item
$a\BE\cdots\BE b \iff a\in\district{b}$;
\item
$a\DE c\BE\cdots\BE b\iff \child{a}\in\district{b}$;
\item
$a\BE\cdots\BE c\LDE b\iff a\in\district{\child{b}}$;
\item
$a\DE c\BE\cdots\BE d\LDE b\iff \child{a}\in\district{\child{b}}$.
\end{alphlist}
Therefore two vertices $a\in A$ and $b\in B$ are connected by a pure-collider path if and only if the two sets $\district{a\cup\child{a}}$ and $\district{b\cup\child{b}}$ are not disjoint which is equivalent to $\district{A^*\cup\child{A^*}}\cap\district{B^*\cup\child{B^*}}\neq\varempty$.
\end{proof}

\begin{proof}[Proof of Theorem \ref{theorem}]
By Corollary 1 and Proposition 2 of \citet{koster99} we have
\[
A\sepm B\given C\wrt{G}
\iff A\sepm B\given C\wrt{G_{\ancestor{A\cup B\cup C}}}
\iff A^*\sepm B^*\given C\wrt{G_{\ancestor{A\cup B\cup C}}}
\]
for some disjoint subsets $A^*$ and $B^*$ such that
$A\subseteq A^*$, $B\subseteq B^*$ and
$A^*\cup B^*\cup C=\ancestor{A\cup B\cup C}=M$. Letting $H=G_M$.
we obtain by application of the previous lemma
\[
A^*\sepm B^*\given C\wrt{G_{\ancestor{A\cup B\cup C}}}
\iff\districtin{H}{A^*\cup\childin{H}{A^*}}\cap\districtin{H}{B^*\cup\childin{H}{B^*}}=\varempty,
\]
which proves the equivalence of (i) and (iii).  The equivalence of (i) and (ii) has been proved in \citet{TR03} in the case of acyclic simple graphs; the generalization of the proof to the present case is straightforward.
\end{proof}

For construction of the sets $A^*$ and $B^*$, we set $V^*=\ancestor{A\cup B\cup C}$ and consider the subgraph $G_{V^*}$. In a first step, two vertices $v,w\in V^*$ are connected by an undirected edge $v\UE w$ whenever $v$ and $w$ are connected by a pure-collider path with every intermediate vertex being an element in $C$. (This step can be split in two substeps: first, identifying (in a topological sense) all vertices $c\in C$ that are in the same district of the subgraph $G_C$ and, second, inserting the edge $v\UE w$ whenever one of the edges $v\DE c\LDE w$, $v\BE c\LDE w$, $v\DE c\BE w$,or $v\BE c\BE w$ for some $c\in C$ is in $G_{V^*}$). Next, we drop all arrowheads obtaining an undirected graph $G'$ with vertex set $V^*$. Now, the set $A^*$ can be defined as the set of all vertices $v\in V^*\without(B\cup C)$ that are not separated from $A$ by $C$ (that is, there exists a path from $v$ to $A$ that does not intersect $C$).
Finally $B^*=V^*\without(C\cup A^*)$. It is clear from this construction of $A^*$ and $B^*$ that $A^*$ and $B^*$ are $m$-separated given $C$ if and only if $A^*$ and $B^*$ are not adjacent in the undirected graph $G'$ if and only if property (iii) of Theorem \ref{theorem} holds.

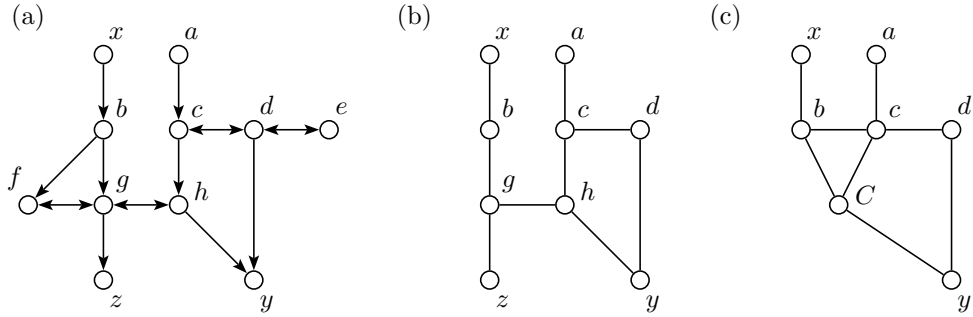
\begin{figure}
\begin{center}
\begin{pspicture}(0,0)(5,4.5)
  \footnotesize
  \psset{linewidth=0.6pt,arrowscale=1.5 2.0,arrowinset=0.2,dotsep=2pt}  
  \cnode(1.5, 3.5){0.13}{1}\nput[labelsep=3pt]{060}{1}{$x$}
  \cnode(1.5, 2.5){0.13}{2}\nput[labelsep=3pt]{045}{2}{$b$}
  \cnode(1.5, 1.5){0.13}{3}\nput[labelsep=3pt]{045}{3}{$g$}
  \cnode(1.5, 0.5){0.13}{4}\nput[labelsep=3pt]{300}{4}{$z$}
  \cnode(0.5, 1.5){0.13}{5}\nput[labelsep=3pt]{120}{5}{$f$}
  \cnode(2.5, 3.5){0.13}{6}\nput[labelsep=3pt]{060}{6}{$a$}
  \cnode(2.5, 2.5){0.13}{7}\nput[labelsep=3pt]{045}{7}{$c$}
  \cnode(2.5, 1.5){0.13}{8}\nput[labelsep=3pt]{030}{8}{$h$}
  \cnode(3.5, 2.5){0.13}{9}\nput[labelsep=3pt]{060}{9}{$d$}
  \cnode(3.5, 0.5){0.13}{10}\nput[labelsep=3pt]{300}{10}{$y$}
  \cnode(4.5, 2.5){0.13}{11}\nput[labelsep=3pt]{060}{11}{$e$}
  \ncline{->}{1}{2}
  \ncline{->}{2}{3}
  \ncline{->}{3}{4}
  \ncline{->}{2}{5}
  \ncline{->}{6}{7}
  \ncline{->}{7}{8}
  \ncline{->}{8}{10}
  \ncline{->}{9}{10}
  \ncline{<->}{3}{5}
  \ncline{<->}{3}{8}
  \ncline{<->}{7}{9}
  \ncline{<->}{9}{11}
  \rput(0.5,4){(a)}
\end{pspicture}
\begin{pspicture}(0,0)(4,4.5)
  \footnotesize
  \psset{linewidth=0.6pt,arrowscale=1.5 2.0,arrowinset=0.2,dotsep=2pt}  
  \cnode(1.5, 3.5){0.13}{1}\nput[labelsep=3pt]{060}{1}{$x$}
  \cnode(1.5, 2.5){0.13}{2}\nput[labelsep=3pt]{045}{2}{$b$}
  \cnode(1.5, 1.5){0.13}{3}\nput[labelsep=3pt]{045}{3}{$g$}
  \cnode(1.5, 0.5){0.13}{4}\nput[labelsep=3pt]{300}{4}{$z$}
  \cnode(2.5, 3.5){0.13}{6}\nput[labelsep=3pt]{060}{6}{$a$}
  \cnode(2.5, 2.5){0.13}{7}\nput[labelsep=3pt]{045}{7}{$c$}
  \cnode(2.5, 1.5){0.13}{8}\nput[labelsep=3pt]{030}{8}{$h$}
  \cnode(3.5, 2.5){0.13}{9}\nput[labelsep=3pt]{060}{9}{$d$}
  \cnode(3.5, 0.5){0.13}{10}\nput[labelsep=3pt]{300}{10}{$y$}
  \ncline{-}{1}{2}
  \ncline{-}{2}{3}
  \ncline{-}{3}{4}
  \ncline{-}{6}{7}
  \ncline{-}{7}{8}
  \ncline{-}{8}{10}
  \ncline{-}{9}{10}
  \ncline{-}{3}{8}
  \ncline{-}{7}{9}
  \rput(0.5,4){(b)}
\end{pspicture}
\begin{pspicture}(0,0)(4,4.5)
  \footnotesize
  \psset{linewidth=0.6pt,arrowscale=1.5 2.0,arrowinset=0.2,dotsep=2pt}  
  \cnode(1.5, 3.5){0.13}{1}\nput[labelsep=3pt]{060}{1}{$x$}
  \cnode(1.5, 2.5){0.13}{2}\nput[labelsep=3pt]{045}{2}{$b$}
  \cnode(2.0, 1.5){0.13}{3}\nput[labelsep=3pt]{020}{3}{$C$}
  \cnode(2.5, 3.5){0.13}{6}\nput[labelsep=3pt]{060}{6}{$a$}
  \cnode(2.5, 2.5){0.13}{7}\nput[labelsep=3pt]{045}{7}{$c$}
  \cnode(3.5, 2.5){0.13}{9}\nput[labelsep=3pt]{060}{9}{$d$}
  \cnode(3.5, 0.5){0.13}{10}\nput[labelsep=3pt]{300}{10}{$y$}
  \ncline{-}{1}{2}
  \ncline{-}{2}{3}
  \ncline{-}{2}{7}
  \ncline{-}{6}{7}
  \ncline{-}{7}{3}
  \ncline{-}{3}{10}
  \ncline{-}{9}{10}
  \ncline{-}{7}{9}
  \rput(0.5,4){(c)}
\end{pspicture}
\end{center}

\caption{Example for separation criterion: (a) mixed graph; (b) undirected graph $G'$ over $\ancestor{\{x,y,z\}}$; (b) undirected graph $G'$ over $\ancestor{\{x,y,g,h\}}$}
\label{examplegraph}
\end{figure}

\begin{example}
We illustrate the separation criterion by the graph depicted in Figure \ref{examplegraph}(a) taken from Figure 2 of \citet{TR03}. Suppose that we are interested whether $x$ and $y$ are separated by $z$. We follow the above construction of the graph $G'$. For the first step, nothing is to do as the vertex $z$ is only connected by a single edge $g\DE z$. Thus, deleting vertices $f$ and $e$ as they do not belong the the ancestral set $\ancestor{\{x,y,z\}}$, and omitting all arrowheads, we obtain the undirected graph $G'$ in Figure \ref{examplegraph}(b). This graph contains the path $x\UE b\UE g\UE h\UE y$ between $x$ and $y$ not intersecting $z$, which implies that sets $A^*$ and $B^*$ of the desired from cannot be found and hence that $x$ and $y$ are not $m$-separated given $z$.

We note that subpaths of the form $g\DE z\LDE g$ do not lead to insertion of self-loops $g\UE g$ as such self-loops are irrelevant for separation in the finally obtained undirected graph $G'$.

For a slightly more complicated example, let $C=\{g,h\}$. To see whether $x$ and $y$ are $m$-separated given $C$, we first identify  the two vertices $g$ and $h$ as they are in the same district. Next, we add an edge $b\UE c$ because of the path $b\DE C\LDE c$. Removing all arrowheads, we obtain the graph in Figure \ref{examplegraph}(c), which shows that $x$ and $y$ are not $m$-separated given $C$.
\end{example}

\bibliography{papers,application}
\bibliographystyle{stat}

\end{document}